\documentclass[preprint]{sig-alternate}
\usepackage[T1]{fontenc} 
\usepackage[utf8]{inputenc}
\usepackage[table]{xcolor}

 \usepackage{array,multirow,graphicx}
\usepackage{tabularx}
\usepackage{balance}
\usepackage{pdflscape}
\usepackage{float}
\usepackage{xspace} 
\usepackage{hyperref}
\usepackage{listings}
\usepackage{framed}
\usepackage{amsmath}
\usepackage{amssymb}
\usepackage{datatool}
\usepackage{textpos}
\usepackage[linesnumbered]{algorithm2e}
\usepackage{numprint}
\usepackage{listings}

\newtheorem{definition}{Definition}

\begin{document}

\newcommand{\FIXME}[1]{{\color{red}#1}}
\newcommand{\nbprograms}{\numprint{9}\xspace}
\newcommand{\nbsosies}{\numprint{30184}\xspace}
\newcommand{\nbsosiesMR}{\numprint{12291s}\xspace}
\newcommand{\transfo}[1]{``#1''}
\newcommand{\checkablenumber}[1]{\FIXME{#1}}

\lstset{
  basicstyle=\ttfamily\scriptsize,        
 keywordstyle=\ttfamily\bf,
stringstyle=\color{red}\ttfamily,
commentstyle=\color{blue}\ttfamily,
  breakatwhitespace=false,         
  breaklines=true,                 
  captionpos=b,                    
  extendedchars=true,              
  language=Java,                 
  showspaces=false,                
  showstringspaces=false,          
  showtabs=false,                  
  tabsize=2                       
}

\title{Tailored Source Code Transformations to Synthesize Computationally Diverse Program Variants}

\numberofauthors{3}
\author{
\alignauthor
Benoit Baudry\\
\affaddr{INRIA/IRISA}\\
       \affaddr{Rennes, France}\\
\email{\small benoit.baudry@inria.fr}
\alignauthor
Simon Allier\\
\affaddr{INRIA/IRISA}\\
\affaddr{Rennes, France}\\
\email{\small simon.allier@inria.fr}
\alignauthor
Martin Monperrus\\
\affaddr{University of Lille \& INRIA}\\
\affaddr{Lille, France}\\
\email{\small martin.monperrus@univ-lille1.fr}
} 

\toappear{\emph{Acknowledgements:}
We thank Ioannis Kavvouras for his participation to the experimentation, Westley Weimer and Eric Schulte for their expert feedback on this paper, as well as our colleagues for insightful discussions and feedback.
This work is partially supported by the EU FP7-ICT-2011-9 No. 600654 DIVERSIFY project.}

\maketitle

\begin{textblock*}{20cm}(-1cm,-1.3cm)
\begin{center}
Technical Report, Inria, 2013
\end{center}
\end{textblock*}

\begin{abstract}
The predictability of program execution provides attackers a rich source of knowledge who can exploit it to spy or remotely control the program. 
Moving target defense addresses this issue by constantly switching between many diverse variants of a program, which reduces the certainty that an attacker can have about the program execution. The effectiveness of this approach relies on the availability of a large number of software variants that exhibit different executions. However, current  approaches rely on the natural diversity provided by off-the-shelf components, which is very limited. 
In this paper, we explore the automatic synthesis of large sets of program variants, called \textit{sosies}.
Sosies provide the same expected functionality as the original program, while exhibiting different executions. They are said to be computationally diverse.

This work addresses two objectives: comparing different transformations for increasing the likelihood of sosie synthesis (densifying the search space for sosies); demonstrating computation diversity in synthesized sosies.
We synthesized \nbsosies sosies in total, for \nbprograms large, real-world, open source applications. For all these programs we identified one type of program analysis that systematically increases the density of sosies; we measured computation diversity for sosies of 3 programs and found diversity in method calls or data in more than 40\% of sosies. This is a step towards controlled massive unpredictability of software.
\end{abstract}

\section{Introduction}
\label{sec:intro}

Predictability of software execution is a weakness with respect to cybersecurity. 
For example, the ability to predict a program's memory layout or the set of machine code instructions allows attackers to design code injection attacks. 
All solutions that address the mitigation of these weaknessses are founded on the diversification of programs or their environments. 
For example, address space layout randomization \cite{shacham04} introduces artificial diversity by randomizing the memory location of certain system components. The objective is to make the memory layout unpredictable from one machine to another, or even from one run of the program to another \cite{xu03}. 
Similarly, instruction set randomization \cite{kc03} generates a diversity of machine instructions to prevent the predictability of the assembly language for a given architecture. 

More recently, moving target defense proposes to use a large number of program variants and to continually shift between them at runtime \cite{jajodia11}. This approach aims at making the attack space unpredictable to the attacker by reducing the predictability about a program's control or data-flow. The success of moving target defense relies on two essential ingredients: the availability of a large number of program variants that implement diverse executions, and; the ability of switching between variants at runtime. 
This work focuses on the first ingredient. 
\emph{We propose a novel technique to automatically synthesize a large set of program variants that provide the same expected functionality as the original program and yet exhibit computation diversity.}

We define a novel form of program variant that we call \textit{sosie programs} (``sosie'' is  French for ``look-alike''). P' is said to be a sosie of a program P if the code of P' is different from P and P' still exhibits the same verified external behavior as P, \textit{i.e.}, still passes the same test suite as P. 
This work compares different program transformations for the automatic synthesis of sosie programs.
The process consisting of searching for program variants satisfying our sosie definition is called ``sosiefication'', the set of all possible program variants obtainable with the transformations forms the search space of sosie synthesis.

All the considered transformations have a random component to explore the space of all program variants. Yet, from an engineering perspective, randomness does not mean inefficient, and we want these transformations to produce large quantities of sosies in a reasonable amount of time. 
The goal of the transformations is thus to increase the likelihood of sosie synthesis, given a fixed budget (\textit{e.g.} time or resources). 
\emph{Consequently, we compare different kinds of program analysis and transformations with respect to their ability of confining the search in a space in which the density of potential sosies is high.}

Also,  with respect to moving target defense, the resulting sosies must exhibit  executions different from the one of the original program. We measure this diversity in terms of divergence in method calls and data between sosies and the original.

We present an extensive evaluation of the synthesis of sosie programs.
We set up 9 program transformations, some of them being purely random while others involve some program analysis. They are all based on the same idea
of removing, adding or replacing statements in source code. 
The transformations are applicable to Java programs and applied on \nbprograms open source code bases.
This enables us to answer two main research questions:
1) what are the most fruitful synthesis techniques to generate sosie programs (sosies density)?
2) how is the execution of sosies different from the execution of the original program (computational diversity)?

To sum up, the contributions of the paper are:
\begin{itemize}
 \itemsep0em 
\item the definition of ``sosie program'' and ``sosiefication'';
\item 9 source code transformations for the automatic synthesis of sosie programs;
\item the empirical evidence of the existence of very large quantities of software sosies given our transformations and dataset in Java;
\item the empirical evaluation of the effectiveness of those different transformations with respect to sosies density and computational diversity.
\end{itemize}

The paper is organized as follows.
Section \ref{sec:sosification} defines the concepts of ``sosie software'' and ``sosiefication''. 
Section \ref{sec:evaluation} presents a large scale empirical study on the presence of software sosies and the difficulty of synthesizing them.
Section \ref{sec:rw} outlines the related work and section \ref{sec:conclusion} sets up a research agenda on the exploitation of computational diversity.

\section{Software Sosies}
\label{sec:sosification}

In this section we define what a ``software sosie'' is.
We discuss an automatic synthesis process of software sosies based on source code transformation and static analysis.  
We describe how the process can be configured with different transformation strategies.

\subsection{Definition of Software Sosie} 
\label{sec:def}

\begin{definition}\label{def:sosie} \textbf{Sosie} (noun).
  Given a program $P$, a test suite $TS$ for $P$ and a program transformation $T$, a variant $P'$=$T(P)$ is a sosie of $P$ if the two following conditions hold
1) there is at least one test case in $TS$ that executes the part of $P$ that is modified by $T$
2) all test cases in $TS$ pass on $P'$.
\end{definition}

Sosies are identical to the original program with respect to the test suite: they have the same oberved behavior as $P$. 
The word sosie is a French word that literally means ``look alike'': there exists ``sosies'' of Madonna (the famous singer).
Since software sosies do not have a visual component, we propose the term ``sosie'' as an alternative to ``look alike''.
From a behavioral perspective, the sosies of $P$ ``look like'' $P$, since they exhibit the same observable behavior.
\emph{The objective of sosies is to provide behaviorally identical yet computationally diverse variants of a program.}

\begin{definition}\label{def:sosie} \textbf{Sosifiecation}.
Sosiefication is the process of synthesizing software sosies. Sosie synthesis is performed through source code transformation on a program P and produces program variants, some of being sosies. 
\end{definition}

The ultimate  transformation for sosie synthesis would ensure for sure that 
1) the resulting program will be a sosie and 
2) the resulting program will be computationally diverse (i.e. its execution would be different w.r.t. to a domain specific monitoring security criteria).
This is a hard problem, instead of transformations that would yield 100\% of sosies, we study transformations that maximize the likelihood of finding interesting sosies.

Consequently, a sosie synthesis is not performed through ``any'' code transformation. The transformation is carefully crafted with a clear objective in mind. It  is not a random mutation but the voluntary modification of one piece of code.
In this paper, we discuss sosiefication in general and \emph{9 sosie synthesis transformations, which all have the
objective of maximizing the likelihood of finding interesting sosies.}

If sosies can appear as mutants in the sense of mutation testing \cite{demillo78}, we believe they are fundamentally different. The intention of the process is different: sosie synthesis aims at generating software diversity meant to be used in production while mutants for mutation testing are meant to simulate faults in order to improve the fault detection power of test suites. The intention of the transformations is different: mutation operators are meant to mimic faults while our transformations are meant to increase computation diversity while keeping the same observable behavior.

\subsection{Synthesis of Software Sosies} 

\begin{figure}
  \centering
  \includegraphics[width=\columnwidth]{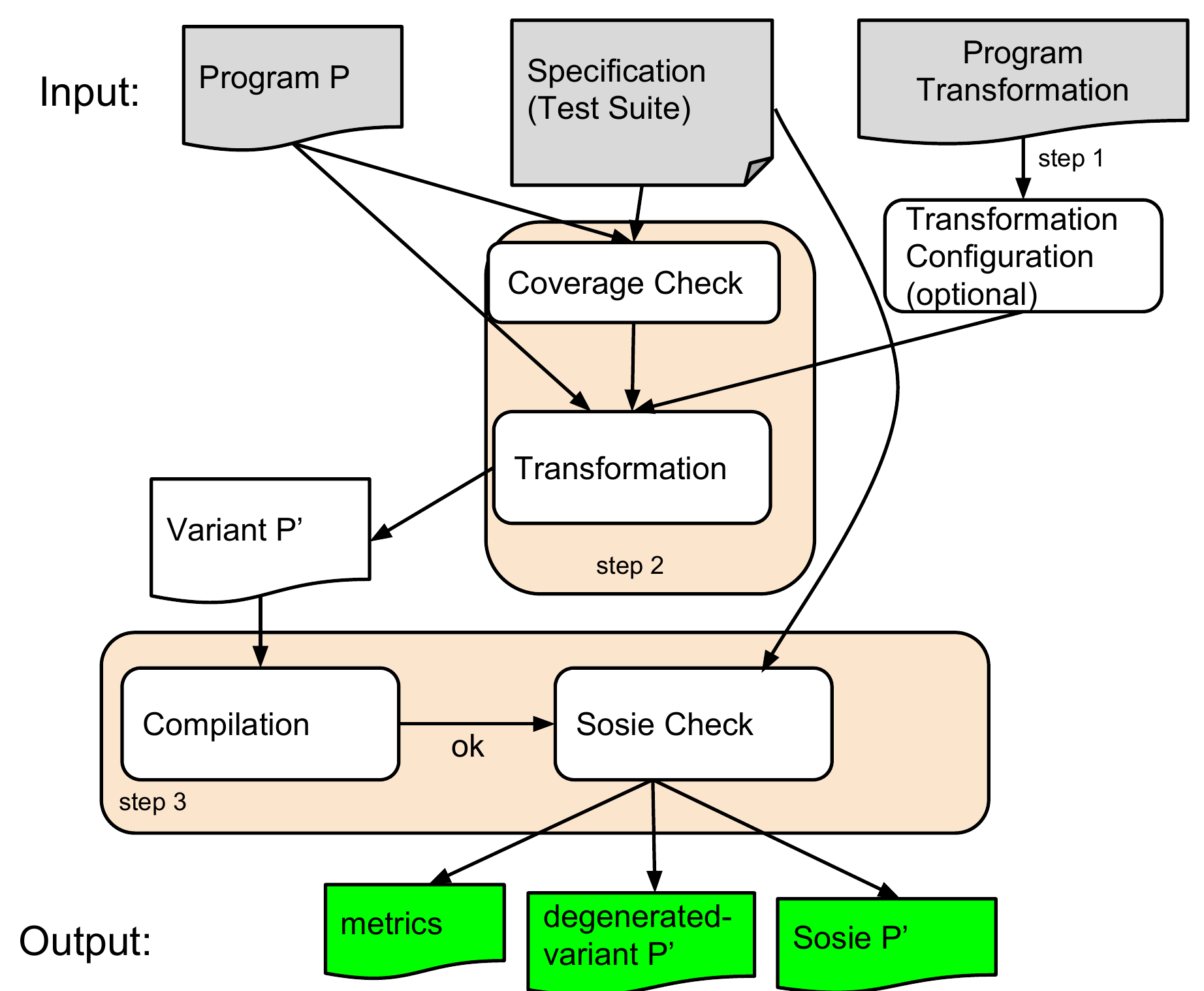}
  \caption{Sosiesifcation is the process of synthesizing software sosies.}
  \label{fig:sosie_flow}
\vspace{-.2cm}
\end{figure}

The sosiefication (sosie synthesis) process takes three kinds of input: a program for which one wants to generate sosies, the test suite for this program, and a  program transformation.
The transformation can optionally be configured or calibrated for the software under sosiefication.
Then, the transformation is applied to generate as many program variants as needed. 
Those program variants are candidate to be sosies.
The variants are executed against the test suite to assert whether they actually do what they preserve the original functionality defined by the test suite.
If the test suite passes, they are real sosies. 
Figure \ref{fig:sosie_flow} illustrates this synthesis process.

\subsection{Program Transformations for Sosiefication} 

We propose source code transformations that are based on the modification of the abstract syntax tree (AST).
As previous work \cite{legoues12,Schulte13}, we consider three
families of transformation that manipulate statement nodes of the AST:
1) remove a node in the AST (Delete);
2) adds a node just after another one (Add);
3) replaces a node by another one, e.g. a statement node is replaced by another statement (Replace). 

For ``Add'' and ``Replace'',  the \textbf{transplantation point} refers to where a statement is inserted, the \textbf{transplant statement} refers to the statement that is copied and inserted and both transplantation and transplant points are in the same AST (we do not synthesize new code, nor take code from other programs).
For ``Add'', the transplantation point is a location between two existing statements, for ``Replace'', it refers to the \textbf{replacee}, the statement that is replaced. The set of all statements that can transplanted at a given point are called \textbf{transplant candidates}.

The transformations consisting of adding, replacing and deleting, random are called \transfo{add-Random}, \transfo{replace-Random} and \transfo{delete}. Those transformations provide us with a baseline to analyze the efficiency of sosiefication.

Despite the simplicity of the delete/add/replace, it is still possible to perform several analyses to increase the likelihood of finding sosies within a given sample of variants.
First, one can add preconditions on the statements to be added or replaced (Section \ref{sec:preconditions}).
Second, one can exploit the fact that names have sense (Section \ref{sec:wittgenstein}).
Third, one drive the addition and replacement with the information given by static type declarations (Section \ref{sec:reactions}).
All those analysis express a kind of compatibility between the transplantation point and the transplant statement.
In total, we will define nine source code transformations.

\subsection{Preconditions for ``Add'' and ``Replace''} 
\label{sec:preconditions}

There are different reasons for which a random add and replace fails at producing a compilable variant. In Java, for instance, the control flow must remain consistent (if not declared as returning ``void'', a method has to contain a return statement in all branches).
Hence we introduce different preconditions to limit the number of meaningless variants.
First,  the \texttt{delete} transformation never removes control flow AST nodes (e.g. return statements).
Also, for \texttt{replace} and \texttt{add}, we enforce that: a statement cannot be replaced by itself; statements of type \emph{case}, AST nodes of type \emph{variable instantiation}, \emph{return}, \emph{throw} are only replaced by statements of the same type; the type of returned value in a \emph{return} statement must be the same for the original and new statement.

\subsection{Name-driven Sosie Synthesis}
\label{sec:wittgenstein}

Program names are not random. They carry some meaning, some intention. 
H{\o}st and {\O}stvold have even shown that one can analyze programs based on the names that are used \cite{Host2009}.
In the context of sosiefication, our intuition is that if one adds or replaces snippets that refer to similar identifiers, they are likely to manipulate the same kinds of objects, both in terms of syntax and semantics. 

Hence we define the two following transformations for sosie synthesis:
\begin{description}
  \item[add-Wittgenstein] adds an AST node that refers to variables names that are manipulated in the statement that just precedes the transplantation point\footnote{consequently, one never inserts a statement at the very beginning of a block}.  
  \item[replace-Wittgenstein] replaces an AST node with a new one that refers to variables names that are used in the replacee. 
\end{description}
In the spirit H{\o}st and {\O}stvold, their names refer to the philosopher Ludwig Wittgenstein and his idea that ``meaning is use''. In a programming language context, this could be translated as names carry the type and even more, the domain semantics. By matching names, it is likely to transplant statements that manipulate close concepts. 

\subsection{Static Types for Sosie Synthesis}
\label{sec:reactions}

``Add'' and ``Replace'' manipulate statements that refer to variables. In a programming language with static typing, 
1) those variables must be declared somewhere in the current scope and 2) the expressions assigned to those variables must be consistent with the declared variable type.

We propose to use the static typing information for driving the sosiefication transformations.
The idea is that the transplant statements must refer to types for which there exists variables at the scope of the transplantation points.

This is done in two phases.
First a pre-processing step collects the types of variables for all statements of the program. Second, at transplantation time, the typing precondition is checked. With the former, we collect a set of program specific ``reactions''. 
 
\begin{definition}\label{def:reaction} \textbf{Reaction}.
 A reaction characterizes a code snippet at a certain granularity in the AST (expression, statement, block).
A reaction is a tuple formed of:
1) the list of all variable types that are used in the snippet, this is the  \textbf{input context}
2) the return type of the statement (or ``void'' if not), this is the snippet's \textbf{output context}.
\end{definition}

At transplantation time, we draw in the set of all reactions those that are compatible with a transplantation point.
Following up the biological metaphor of transplantation,  
the choice of the term \textit{reaction} is made in reference to the reactions that drive the cell metabolism \footnote{http://www.nature.com/scitable/topicpage/cell-metabolism-14026182}. In our case, they drive the sosiefication.

For example, for the snippet \texttt{bar(varA , 10 + i);}, the corresponding reaction is: {input context: \texttt{[StaticType(VarA), int]}; assuming that method \texttt{bar} returns a boolean, then the
 output context is: \texttt{boolean}.

To use reactions in the code transformations, we first collect the reactions for each node in the AST.
Then, to ``Add'' or ``Replace'' a node at transplantation point $tp$, we look for a compatible transplant, \textit{i.e.} a reaction which input context contains only types that are in the input context of $tp$, and same thing  for the output context. ``Add'' transformation then adds the transplant and keeps $tp$, while ``Replace''adds the transplant and removes $tp$.

To sum up:
\begin{description}
  \item[add-Reaction] adds an AST node that is type-compatible with the types of variables that are manipulated in the statement that just precedes the transplantation point.  
  \item[replace-Reaction] replaces an AST node with a new one that is type-compatible with the variables names that are used in the replacee. 
\end{description}

Once a transplantation's reaction matches, it happens that the variable names mismatch. For this reason, we add a last step before transplantation:
if two variables (one from the transplantation context and one that is used in the transplant) are compatible, we rename the variable reference of the transplant to the name defined in the transplantation context (if there are several possibilities, we pick randomly one).
This is the essence of the two last sosiefication transformations \transfo{add-Steroid} and \transfo{replace-Steroid}. Compared to \transfo{add-Reaction} and \transfo{replace-Reaction}, they add variable renaming. They are ``on steroids'' in the sense that they give the best empirical results, in particular the fastest sosiefication speed (as shown in Section \ref{sec:evaluation}). 
\begin{description}
  \item[add-Steroid] adds an AST node that is type-compatible with the variables names and bound to existing variables of the transplantation point.
  \item[replace-Steroid] replaces an AST node with a new one that is type-compatible with the variables names and bound to existing variables of the transplantation point.
\end{description}

To recap, we define nine sosiefication transformations:
delete, 
add-Random, replace-Random,
add-Wittgenstein, replace-Wittgenstein,
add-Reaction, replace-Reaction,
add-Steroid, replace-Steroid.

They are not mutually exclusive, some pairs may overlap for some transplantation points and transplant candidates.
Hence, there already is a small probability that two different transformations produce the very same sosie.

\subsection{Additional Checks}

It is meaningless to modify code that is not executed. The resulting variants would be trivially sosies. Hence, we always check that the transplantation point is covered by the test suite, using the Jacoco library\footnote{\url{http://www.eclemma.org/jacoco/}}. 

Also, we aim at comparing the efficiency of the different transformations. 
Hence, we only consider transplantation points for which there is, at least, one compatible reaction for transplant (a reaction which input and output contexts match the ones of the transplantation point). 

\subsection{Outcome of the Sosie Synthesis Process}

The very last check of sosie synthesis consists in checking whether the variant program actually is a sosie. 
This is done as follows. First, we try to compile the variant AST. If the variant compiles, we then run all test cases in $TS$ on the variant. If all test cases pass, the variant is a sosie (according to our definition of sosie), otherwise we call it a degenerated variant, and we throw it away.

To sum up, the output of the process is as follows.
First, it gives a set of transformed programs partitioned in three sets $A$, $B$, and $C$: the \textbf{sosies} $A$ and the \textbf{degenerated-variants}, i.e., the set $B$ of variants that compile but for which at least one test case in the suite has failed, the set $C$ of ill-formed variants (do not compile).
Second, we compute a variety of metrics for understanding the nature of sosiefication and evaluating the transformations: 
the number of transplantation points on which a transformation  has been applied;
the number of \textbf{variants} ($|A|+|B|+|C|$); 
the number of transplant candidates per transplantation points.

It is possible to apply a sosiefication transformation only once or several times in a row. 
In this paper we  focus on the former, first-order sosiefication (only one transformation is applied only once). This simplification enables us to understand the foundations of sosiefication. Note that even if there is only one transformation, it can be large: for ``Add'' and ``Replace'' transformations, the transplant can be large (a large statements or several statements combined in a block),  for ``Delete'' and ``Replace'' transformations, the removed code can also be large, up to a full block (a block is a statement).

\textbf{Prototype Implementation:} We use the Spoon framework \cite{spoon} to parse and transform Java abstract syntax trees. We focus on Java projects that use JUnit as a testing framework. 

Listing \ref{ex:sosie} displays an example of sosie in commons.collections synthesized with the ``Replace-Steroid'' transformation. The return statement has the following input context: \texttt{[SortedMap<K, V> tmpMap, Predicate keyPredicate, Predicate valuePredicate] } and an empty  output context. It is replaced by another return statement, which input context is \texttt{[SortedMap map]} and output context is also empty. The variable in the transplant return statement is mapped to \texttt{tmpMap}. 
\begin{lstlisting}[caption={A real world sosie created in Apache Commons Collections },label=ex:sosie,float]
public SortedMap<K, V> headMap(final K toKey) {
  final SortedMap<K, V> tmpMap = getSortedMap().headMap(toKey);
  return new PredicatedSortedMap<K, V>(tmpMap, keyPredicate, valuePredicate);
   // replaced by return new UnmodifiableSortedMap<K, V>(tmpMap);
}
\end{lstlisting}
\label{example:replace}

\section{Empirical Inquiry on Sosiefication}
\label{sec:evaluation}

\begin{table*}[ht]
\centering
\begin{tabularx}{\textwidth}{lXXXXXXXXX}
\hline
& \#LoC & \#classes& \#test cases& \#assert  &  coverage & \#stmt & \#transf.  stmt & compile time (sec)& test time (sec)\\
\hline
JUnit  & 8056  & 170  & 721  & 1535 & 82\% & 2914 & 1654 & 4.5 & 14.4\\
EasyMock& 4544  & 81 & 617  & 924& 91\% & 2042 & 1441 & 4 & 7.8 \\
Dagger (core)& 1485  & 23 & 128  & 210 &  85\%& 674 & 95 &5.1 & 11.2 \\
JBehave-core & 13173  &  188 & 485  & 1451 & 89\% & 4984 & 3405 & 5.5 & 22.9\\
Metrics & 4066 & 56 & 214  & 312 & 79\%  & 1471 & 319 &4.7 & 7.7 \\
commons-collections & 23559 & 285 & 1121  & 5397& 84\% &9893 & 5027 & 7.9& 22.9\\
commons-lang & 22521 & 112  & 2359  & 13681 & 94\% & 11715 & 9748 & 6.3 & 24.6\\
commons-math & 84282 &  803 & 3544 & 9559 & 92\% & 47065 &12966 & 9.2& 144.2\\ 
clojure & 36615 & 150 & N\/A  &N\/A & 71\% & 18533 &12259 &105.1 & 185\\
\hline
\end{tabularx}
\caption{Descriptive statistics about our experimental data set}
\label{tab:data}
\vspace{-.3cm}
\end{table*}

We now present our experiments on sosies.
\textbf{Our main objective is to gather  knowledge on the sosiefication process}.
The existing body of knowledge on software mutation is biased against sosies. In particular,
previous works have neither tried to maximize the number of sosies nor to evaluate the computational difference between variants.
To our knowledge, only Schulte et al. have studied sosiefication, in the context of C code \cite{Schulte13}.
It is an open question to know whether Schulte et al.'s findings apply to our transformations on object-oriented Java programs (see Sec \ref{rq1}).

\subsection{Analysis Criteria}
In essence, sosiefication is a search problem, where the search space is the set of all possible program variants obtainable with a given transformation.
Hence, sosiefication consists of navigating the search space of program variants, looking for the ones that are identical with respect to the test suite. The navigation is done through small steps, where a step is the application of a code transformation. The application of a transformation is more or less likely to produce sosies as mentioned in Section \ref{sec:def}.
Put another way, transformation rules slice the global search space of variants, to define the search space of a given transformation.
If the resulting space is dense in terms of sosies, it means that the transformation is often successful in finding sosies.
On the other hand, if the space sparsely contains sosie, applying the transformation would rarely yield sosies.

What is really costly in navigating the sosiefication search space is the time to compile the program and even worse, the time to run the test suite (as shown in Table \ref{tab:data}). Hence,  navigation cost dominated by  checking whether the variant is actually a sosie.  This is similar to what happens for program repair as shown by Weimer et al. \cite{Weimer2013}.
Consequently, if the search space defined by a transformation is dense in sosies, one decreases the global time spent in assessing degenerated-variants.

This point may seem theoretical but it has a very practical application. From an engineering persepective, one always want to do the maximum for a given budget. For an engineering team setting up moving target defense, the goal is to find as many sosies as possible within a given time or budget of computation resources. If the team can have 1000 sosies instead of 500 that would be better. This is equivalent to synthesizing space that are dense in sosies. 
\emph{Our first objective is to identify sosie synthesis transformations defining a search space that is dense in sosies.}

For moving target defense, what matters is to create an execution profile that is as much unpredictable as possible.  This requires engineering  variants of source code that are identical to the original, with respect to the observed behavior, but that also produce  executions that are different from the original. \emph{The second objective of this evaluation is to assess whether the synthesized sosies are computationally diverse. }
\\

We can sum our research questions as:
\\

\textbf{RQ1. Do previous results on creating sosies in imperative 
programs hold for our new tailored transformations on object-oriented 
programs?} We replicate the same kind of experiment as Schulte et al. \cite{Schulte13}, but we change the transformations and the dataset (different programs, different programming language) (see Section \ref{rq1}).
\\

\textbf{RQ2. What are the best sosiefication transformations with respect to the density of sosies?} The baseline here is \transfo{add-Random}, \transfo{replace-Random} and \transfo{delete} (see Section \ref{rq2}).
\\

\textbf{RQ3. Are sosies computationally diverse, i.e. do they exhibit executions that are different from the original program?} This requires having a definition of ``computationally diverse'', we present two of them in Section \ref{sec:protocol}.

\subsection{Experimental Design}
\label{sec:exp-design}

\subsubsection{Dataset}
We sosiefy  \nbprograms widely-used open source projects \footnote{Sources of all programs used in our experiments are available here: \url{http://diversify-project.eu/sosiefied-programs/}}. 
The inclusion criteria are that:
1) they come with a good test suite according to the statement coverage ($>70\%$)
2) they are written in Java and are correctly handled by the source code analysis and transformation library we use (Spoon \cite{spoon}).
All test suites are implemented in JUnit, except in the case of Clojure. Clojure is a Lisp-like interpreter, thus the test suite is a set of Lisp programs. 

Table \ref{tab:data} gives the essential metrics on this dataset. The programs range from 1 to 80 KLOC. Given the critical role of test suites in the sosiefication process, we provide a few statistics about the test suites of each program. 
All test suites have a high statement coverage.
To us, test suites with many assertions and high coverage indicate an important effort and care put into their design. 

Table \ref{tab:data} also provides the number of statements for each program. Since all our sosiefication transformations manipulate statement for transplantation, this number is an indicator of the size of the search space for sosiefication. We also provide the time (in seconds) to compile the program and run the test suites. It is important since the time of sosiefication is dominated by the time to compile variants and check they are sosies (by running the test suite). The times are computed on the same machine, \footnote{CPU: Intel Xeon Processor W3540 (4 core, 2.93 GHz), RAM: 6GB} in the same idle state, using an unmodified version of the program and test suite. 

\subsubsection{Protocol}
\label{sec:protocol}

\paragraph{Sosiefication}

The experimental protocol is described in Algorithm \ref{batch_sosie}.
For each program of our dataset, as long as we have available computing power, we draw a transplantation point and run the 9 sosiefication transformations on it. 
This protocol is budget based: we try neither to exhaustively visit the search space nor to have a fixed sample. 
Our computation platform is Grid5000, a scientific platform for parallel, large-scale computation \cite{bolze2006grid}.
We submit one batch for each program, they run as long as resources (CPU and memory) are available. 
We look for sosies as long as we have available free computing slots on Grid5000.

This protocol samples the search space of all possible statement transformations at two levels: (1) sample the transplantation points (those statements for which there is at least one compatible reaction, line 4 of Algorithm \ref{batch_sosie}); (2) given a statement selected as a transplantation point, sample the set of transplant candidates (line 8 of Algorithm \ref{batch_sosie}). 

Eventually, we obtain a number of sosies for each software application under study. In addition to that, we know the number of transplantation points that were tried and the number of ill-formed variants that do not compile. By carefully characterizing the two levels of sampling, this enables us to answer our 3 research questions.

\begin{algorithm}[t]
  \DontPrintSemicolon
  \KwData{$P$, a program we want to sosiefy}
  \KwResult{data for table \ref{tab:sosies-global}}
  $S$=\{statements in the AST of P\}\;
  $R$=\{reactions extracted from the $P$\}\;
  \While{resources\_available} 
  {
    randomly select a transplantation point $stmt \in S$ \;
    $Comp_R \leftarrow  \{r \in R$ | r is compatible with stmt\}\; 
    \If{$Comp_R$$\neq\emptyset$}
    {
  \ForEach{t in the 9 transformations}
  {
      \If{$t$ requires a reaction}{select a random one in $Comp_R$}
      variant $\leftarrow$ application of t on $stmt$ \;
      compile t\;
	check if the variant is a sosie\;
      if yes, save it for future analysis\;
 }
    }
  }
  \caption{The experimental protocol for evaluating our 9 sosiefication transformations}
  \label{batch_sosie}
\end{algorithm}

\paragraph{Computation monitoring}

We quantify computation diversity by measuring  \textbf{method calls diversity} and \textbf{variable diversity}. At each method entry, we log the method signature (class name, method signature) to gather one sequence of method calls for each execution. A difference between the sequence of the original program and a sosie indicates method calls diversity. This metric has been shown to be a relevant way of capturing the ``sense of self'' of a program and distinguish it from another implementation by Forrest et al. \cite{forrest96}. The values of all data (variables, parameters, attributes) in the current scope are logged at each control point (conditional, loop). For object variables, we collect their string representation (i.e. the return value of method toString() in Java). A difference between sequence of variable values of the original program and a sosie, indicates variable diversity.
 
Trace comparison is performed as described in algorithm  \ref{trace_compare}. The 'cleaning' step in line 5 of algorithm \ref{trace_compare} looks for data or method calls, which always change from one run to another (\textit{e.g.}, temporary files generating during execution always have a different name) in order to discard them in the comparison.

\begin{algorithm}[t]
  \DontPrintSemicolon
  \KwData{$Pool$ a set of sosies in which we look for computation diversity, $P$ an original program, $TS$ the test suite for $P$}
  \KwResult{data for table table \ref{tab:diffTrace}}
  \ForEach{$sosie\in Pool$}
  {\ForEach{$test\in TS$}
  {
    $Trace_P\leftarrow$ test run on P\;
  $Trace_{sosie}\leftarrow$ test run on sosie\;
  remove 'noisy' data and method calls\;
  compare\_call\_traces($Trace_P$ and $Trace_{sosie}$)\;
  compare\_data\_trace($Trace_P$ and $Trace_{sosie}$)\;
  }
  count the number of test cases for which $s$ exhibits diversity on data or method calls
  }
  \caption{Measurement of computational diversity}
  \label{trace_compare}
\end{algorithm}

\begin{table*} 
\centering
\begin{scriptsize}
\begin{tabular}{crlllllll}
  \hline
 &&    \#tested  & \# candidate & \# tested & \# compilable & \# sosies &sosie & sosies/h  \\ 
 &&   statements & & variants &  & & density &    \\ 

 \hline
\parbox[t]{2mm}{\multirow{9}{*}{\rotatebox[origin=c]{90}{junit}}} &
\textbf{add-Steroid}  & \textbf{669 (40\%)} & \textbf{7554} & \textbf{1246 (17\%) }& \textbf{698 (56\%)} & \textbf{302} & \textbf{24\%} & \textbf{85.5 } \\ 
  &add-Wittgenstein  & 669 (40\%) & 30391 & 865 (3\%) & 279 (32\%) & 114 & 13\% & 59.8  \\ 
  &add-Reaction & 669 (40\%) & 4841 & 1345 (28\%) & 539 (40\%) & 234 & 17\% & 72.1  \\ 
  &add-Rand & 669 (40\%) & 1949466 & 2027 (<1\%) & 231 (11\%) & 154 & 8\% & 46  \\ 
  &\textbf{replace-Steroid}  & \textbf{669 (40\%) }& \textbf{7554 }& \textbf{1270 (17\%)} & \textbf{983 (77\%)} & \textbf{130} &\textbf{10\% }& \textbf{30.1} \\ 
  &replace-Wittgenstein  & 669 (40\%) & 30391 & 876 (3\%) & 307 (35\%) & 82 &9\% & 41.1 \\ 
  &replace-Reaction  & 669 (40\%) & 4841 & 1359 (28\%) & 648 (48\%) & 72 &5\% & 20.3  \\ 
  &replace-Rand  & 669 (40\%) & 1949466 & 2027 (<1\%) & 209 (10\%) & 31 &2\% & 9.4 \\ 
  &delete  & 669 (40\%) & 669 & 669 (100\%) & 423 (63\%) & 63& 9\% & 31.1 \\ 
  
\hline
\parbox[t]{2mm}{\multirow{9}{*}{\rotatebox[origin=c]{90}{easymock}}} &

\textbf{add-Steroid}   & \textbf{943 (65\%)} & \textbf{10456} & \textbf{5246 (50\%)} &\textbf{ 2172 (41\%)} & \textbf{721} & \textbf{14\% }& \textbf{91.9}  \\ 
 & add-Wittgenstein & 943 (65\%) & 52884 & 4951 (9\%) & 1113 (23\%) & 290 & 6\% & 44.5  \\ 
 & add-Reaction  & 943 (65\%) & 9437 & 5197 (55\%) & 1646 (32\%) & 520 & 10\% & 71.3  \\ 
 & add-Rand  & 943 (65\%) & 1925606 & 11284 (<1\%) & 1230 (11\%) & 790 & 7\% & 57.9  \\ 
 & \textbf{replace-Steroid}  & \textbf{943 (65\%)} & \textbf{10460} & \textbf{5633 (54\%) }&\textbf{3823 (68\%) }& \textbf{492} & \textbf{9\%} & \textbf{50.1}  \\ 
 & replace-Wittgenstein  & 943 (65\%) & 52884 & 5059 (10\%) & 1563 (31\%) & 169 & 3\% & 23.9 \\ 
 & replace-Reaction  & 943 (65\%) & 9437 & 5520 (58\%) & 3126 (57\%) & 387 & 7\% & 42.8 \\ 
&  replace-Rand  & 943 (65\%) & 1925606 & 11342 (<1\%) & 1154 (10\%) & 94 & <1\% & 6.9  \\ 
 & delete & 943 (65\%) & 943 & 943 (100\%) & 654 (69\%) & 74 & 8\%& 44.6 \\ 
 \hline 
\parbox[t]{2mm}{\multirow{9}{*}{\rotatebox[origin=c]{90}{dagger}}} &
\textbf{add-Steroid}   & \textbf{85 (90\%)} &\textbf{ 333} &\textbf{ 292 (88\%) }& \textbf{156 (53\%)} & \textbf{20} & \textbf{7\% }&\textbf{29.4} \\ 
&  add-Wittgenstein & 85 (90\%) & 603 & 520 (86\%) & 154 (30\%) & 42 & 8\% & 41.8  \\ 
&  add-Reaction  & 85 (90\%) & 319 & 307 (96\%) & 89 (29\%) & 15 & 5\% & 25.4  \\ 
 & add-Rand & 85 (90\%) & 57290 & 8942 (16\%) & 740 (8\%) & 354 & 4\% & 25.1  \\ 
 & \textbf{replace-Steroid}  & \textbf{85 (90\%)} & \textbf{333} & \textbf{293 (88\%)}& \textbf{208 (71\%)}& \textbf{17} & \textbf{6\% }& \textbf{22.1 } \\ 
 & replace-Wittgenstein  & 85 (90\%) & 603 & 515 (85\%) & 98 (19\%) & 5 & 1\% & 5.5  \\ 
 & replace-Reaction  & 85 (90\%) & 319 & 307 (96\%) & 114 (37\%) & 7 & 2\% & 11.1  \\ 
 & replace-Rand  & 85 (90\%) & 57290 & 8976 (16\%) & 537 (6\%) & 42 & <1\% & 3 \\ 
  &delete  & 85 (90\%) & 85 & 85 (100\%) & 40 (47\%) & 4 & 5\% & 21.2 \\

\hline   
\parbox[t]{2mm}{\multirow{9}{*}{\rotatebox[origin=c]{90}{metrics}}} &
 
\textbf{add-Steroid}   & \textbf{302 (95\%) }& \textbf{1819} & \textbf{1197 (66\%)} & \textbf{702 (59\%) }& \textbf{109} &\textbf{9\%} & \textbf{50.9}  \\ 
&  add-Wittgenstein  & 302 (95\%) & 10553 & 1939 (18\%) & 414 (21\%) & 92 & 5\% & 32  \\ 
&  add-Reaction  & 302 (95\%) & 1297 & 1290 (99\%) & 616 (48\%) & 85 & 7\% & 38.7  \\ 
 & add-Rand & 302 (95\%) & 444242 & 11175 (3\%) & 1471 (13\%) & 782 & 7\% & 49.4  \\ 
 & \textbf{replace-Steroid}  & \textbf{302 (95\%) }& \textbf{1819} & \textbf{1202 (66\%) }& \textbf{850 (71\%) }& \textbf{55}&  \textbf{5\%} & \textbf{24.2} \\ 
 & replace-Wittgenstein & 302 (95\%) & 10553 & 1932 (18\%) & 416 (22\%) & 25 & 1\% & 8.7  \\ 
 & replace-Reaction  & 302 (95\%) & 1297 & 1282 (99\%) & 657 (51\%) & 28 & 2\% & 12.6  \\ 
 & replace-Rand  & 302 (95\%) & 444242 & 11205 (3\%) & 1023 (9\%) & 74 & <1\% & 4.8 \\ 
 &  delete & 302 (95\%) & 302 & 302 (100\%) & 155 (51\%) & 22 & 7\% & 42.1  \\ 
   \hline

\parbox[t]{2mm}{\multirow{9}{*}{\rotatebox[origin=c]{90}{jbehave}}} &

\textbf{add-Steroid}  & \textbf{2798 (82\%)} & \textbf{1852392} & \textbf{12715 (1\%)} & \textbf{6546 (51\%)} & \textbf{3843} & \textbf{30\%} & \textbf{76.5}  \\ 
  &  add-Wittgenstein & 2798 (82\%) & 163225 & 8304 (5\%) & 2922 (35\%) & 1812 & 22\% & 69.5 \\ 
  &  add-Reaction  & 2798 (82\%) & 61594 & 13775 (22\%) & 4839 (35\%) & 3016 & 22\% & 69.8  \\ 
  & add-Rand & 2798 (82\%) & 13945232 & 24917(<1\%) & 1773 (7\%) & 1516 & 6\% & 34.9\\ 
   & \textbf{replace-Steroid} & \textbf{2798 (82\%)} & \textbf{1852392} & \textbf{12625 (1\%)} & \textbf{8293 (66\%)} & \textbf{1818} & \textbf{14\%} & \textbf{30.9} \\ 
  & replace-Wittgenstein & 2798 (82\%) & 163225 & 8277 (5\%) & 2860 (35\%) & 855 & 10\% & 33.2 \\ 
  & replace-Reaction & 2798 (82\%) & 61619 & 13775 (22\%) & 4749 (35\%) & 1088 & 8\% & 25.6  \\ 
  & replace-Rand & 2798 (82\%) & 13945232 & 24904 (<1\%)& 1221 (5\%) & 281 & 1\% & 6.9 \\ 
  & delete & 2798 (82\%) & 2798 & 2798 (100\%) & 1735 (62\%) & 411 & 15\% & 32.8  \\ 
  \hline

\parbox[t]{2mm}{\multirow{9}{*}{\rotatebox[origin=c]{90}{clojure}}} &

\textbf{add-Steroid} & \textbf{1701 (14\%)} & \textbf{13456701703} & \textbf{1677 (<1\%)} & \textbf{609 (36\%)} & \textbf{286} & \textbf{17\%} & \textbf{4.6} \\ 
  & add-Wittgenstein & 1701 (14\%) & 293165 & 1365 (<1\%) & 312 (23\%) & 155 & 11\% & 3.3  \\ 
   & add-Reaction & 1701 (14\%) & 68557 & 1977 (3\%) & 435 (22\%) & 322 & 16\% & 4.8  \\ 
   & add-Rand & 1701 (14\%) & 31524633 & 1925 (<1\%) & 93 (5\%) & 80 & 4\% & 1.4  \\ 
  & \textbf{replace-Steroid} & \textbf{1701 (14\%)} & \textbf{13456701703} & \textbf{1678 (<1\%)} & \textbf{1060 (63\%)} & \textbf{156} & \textbf{9\%} & \textbf{2.2}  \\ 
   & replace-Wittgenstein & 1701 (14\%) & 293165 & 1370 (<1\%) & 395 (29\%) & 103 & 8\% & 2.1  \\ 
  &replace-Reaction & 1701 (14\%) & 68557 & 1930 (3\%) & 515 (27\%) & 83 & 4\% & 1.2  \\ 
  & replace-Rand & 1701 (14\%) & 31524633 & 1907 (<1\%) & 99 (5\%) & 15 & 1\% & 0.3  \\ 
 & delete & 1701 (14\%) & 1701 & 1701 (100\%) & 944 (55\%) & 189 & 11\% & 2.7  \\ 
 \hline

\parbox[t]{2mm}{\multirow{9}{*}{\rotatebox[origin=c]{90}{commons-collections}}} &

\textbf{add-Steroid}  & \textbf{1661 (33\%)} & \textbf{39596} & \textbf{2143 (6\%) }&\textbf{ 645 (30\%)} & 3\textbf{66} & \textbf{17\%} & \textbf{51.7} \\ 
     & add-Wittgenstein  & 1661 (33\%) & 121095 & 2071 (2\%) & 408 (20\%) & 199 & 10\% & 33.5  \\ 
     & add-Reaction & 1661 (33\%) & 32129 & 2459 (8\%) & 461 (19\%) & 318 & 13\%& 45.7 \\ 
     & add-Rand  & 1661 (33\%) & 19458615 & 2885 (<1\%) & 101 (4\%) & 74 & 5\% & 11.7  \\ 
  & \textbf{replace-Steroid} & \textbf{1661 (33\%)} & \textbf{39596} & \textbf{2197 (6\%)} & \textbf{1303 (59\%) }& \textbf{220} & \textbf{10\%} & \textbf{22.1}  \\ 
     & replace-Wittgenstein  & 1661 (33\%) & 121095 & 2112 (2\%) & 608 (29\%) & 148 & 7\% & 21.6 \\ 
    &replace-Reaction & 1661 (33\%) & 32129 & 2504 (8\%) & 786 (31\%) & 101 & 4\% & 12  \\ 
   & replace-Rand & 1661 (33\%) & 19458615 & 2937 (<1\%) & 172 (6\%) & 21 & <1\% & 3.1 \\ 
 & delete& 1661 (33\%) & 1661 & 1661 (100\%) & 973 (59\%) & 174 & 10\% & 23.3  \\ 
\hline
\parbox[t]{2mm}{\multirow{9}{*}{\rotatebox[origin=c]{90}{commons-lang}}} &
\textbf{add-Steroid}   & \textbf{686 (15\%) }& \textbf{50854 }&\textbf{ 7685(<1\%)} & \textbf{2518 (33\%) }& \textbf{1264 } &\textbf{16\%} & \textbf{48.5} \\ 
& add-Wittgenstein  & 686 (15\%) & 72597 & 5593 (8\%) & 1300 (23\%) & 580 & 10\% & 35.6  \\ 
       & add-Reaction  & 686 (15\%) & 17084 & 8338 (49\%) & 1486 (18\%) & 1005 & 12\% & 45.6  \\ 
       & add-Rand& 686 (15\%) & 8036490 & 9957 (<1\%) & 532 (5\%) & 469 & 5\% & 23.4  \\ 
  & \textbf{replace-Steroid} & \textbf{686 (15\%)} & \textbf{50854} & \textbf{7731 (15\%)} & \textbf{4849 (63\%)} & \textbf{756} & \textbf{10\%} & \textbf{20} \\ 
       & replace-Wittgenstein& 686 (15\%) & 72597 & 5590 (8\%) & 1893 (34\%) & 382 & 7\% & 19.8\\ 
      &replace-Reaction & 686 (15\%) & 17084 & 8378 (51\%) & 2612 (31\%) & 340 & 4\% & 12.3  \\ 
     & replace-Rand& 686 (15\%) & 8036490 & 9975 (<1\%) & 535 (5\%) & 50 & 1\% & 2.5  \\ 
   & delete & 686 (15\%) & 686 & 686 (100\%) & 2439 (64\%) & 394  & 10\%& 20.9 \\ 

\hline 
\parbox[t]{2mm}{\multirow{9}{*}{\rotatebox[origin=c]{90}{commons-math}}} &
\textbf{add-Steroid}  & \textbf{1263 (10\%) }& \textbf{8383912305} & \textbf{1076(<1\%)} & \textbf{266 (25\%) }&\textbf{143} &\textbf{13\%} & \textbf{11.2 } \\ 
  & add-Wittgenstein  & 1263 (10\%) & 325832 & 771(<1\%)& 196 (25\%) & 102 & 13\% & 10.9  \\ 
         & add-Reaction & 1263 (10\%) & 1623447 & 1214 (<1\%) & 216 (18\%) & 179 & 15\%& 16 \\ 
      & add-Rand& 1263 (10\%) & 59443095 & 1254 (<1\%)& 65 (5\%) & 49 & 4\% & 8.7  \\ 
  & \textbf{replace-Steroid} & \textbf{1263 (10\%)} & \textbf{8383912305 }& \textbf{1054 (<1\%)} & \textbf{546 (52\%) }& \textbf{83} & \textbf{8\%} & \textbf{3.6 } \\ 
  & replace-Wittgenstein  & 1263 (10\%) & 325832 & 763(<1\%)& 201 (26\%) & 47 & 6\%& 5  \\ 
  &replace-Reaction & 1263 (10\%) & 1623447 & 1226(<1\%) & 138 (11\%) & 20 & 2\% & 2.4  \\ 
  & replace-Rand & 1263 (10\%) & 59443095 & 1256 (<1\%) & 44 (4\%) & 4 & <1\%& 0.8  \\ 
 & delete  & 1263 (10\%) & 1263 & 1263 (100\%) & 726 (57\%) & 115 & 9\% & 3.8 \\

\hline
\end{tabular}
\end{scriptsize}

\caption{Global results for sosie synthesis}
\label{tab:sosies-global}
\end{table*}

\begin{table*}
\centering

\begin{tabularx}{\textwidth}{rXXXXXX}
  \hline
 & \#sosie & diversity & call diversity &  var. diversity & \# of diverse test cases (call div.) & \# of diverse test cases (var. div.) \\ 
  \hline
easymock & 465 & 218 (46.88\%) & 161 (34.62\%) & 139 (29.89\%) & 34.61 & 18.42 \\ 
dagger & 481 & 322 (66.94\%) & 319 (66.32\%) & 19 (3.95\%) & 6.32 & 12.16 \\ 
junit & 446 & 205 (45.96\%) & 194 (43.5\%) & 95 (21.3\%) & 148.86 & 2.32 \\ 
   \hline
\end{tabularx}

\caption{The measuring of computational diversity w.r.t calls and data on a random sample of sosies.}
\label{tab:diffTrace}
\end{table*}

\subsection{Findings}
\label{sec:results}

Table \ref{tab:sosies-global} gives the key metrics of the experiments to answer research questions \#1, \#2, \#3. 
The left-hand side columns of the table give the names of the software application under study and the names of the considered sosiefication transformations.
The column $\#tested\_stmt$ in table \ref{tab:sosies-global} provides data about the first level of sampling: the number of unique statements in the program on which we execute the sosiefication transformations (in parenthesis , the ratio over the total number of candidate statements). 
The column $candidate$ is the sum of transplant candidates over all the tested statements, it is the size of the search space of the second level of sampling mentioned in \ref{sec:protocol}. Its formula is given in footnote\footnote{For \transfo{rand} transformations:\\
$\#candidates=\#tested\_stmt*\#stmt\_in\_prog$

For \transfo{Reaction} transformations:\\
$\#candidates=\sum_{i=1}^{\#tested\_stmt}(\#compatible\_reactions)$

For \transfo{Wittgenstein} transformations:\\
$\#candidates=\sum_{i=1}^{\#tested\_stmt}(\#compatible\_statements)$

For \transfo{Steroid} transformations:\\
$\#candidates=\sum_{i=1}^{\#tested\_stmt}(\#compatible\_reactions \\ * \#variable\_mappings)$

For deletion, the number of candidates is simply $\#tested\_stmt$, since for each tested statement, there was a single candidate for the transformation.
}. 

The column 'variant' is the number of unique actual transformations that have been performed (\textit{e.g.}, if the same transformation is applied twice on the same transplantation point, this counts as one variant). 
The column 'compile' is the number and ratio of variants that compiled, 

The column 'sosies' is the number of variants that are actual sosies. The column 'sosie density' is the ratio of sosies found among all the variants. This sosie ratio found in a random sample of the complete search space (the ratio in the 'variant' column is the proportion of the complete search space actually explored) is an estimate of the sosie density for a given transformation.
The column 'sosies/h' is an approximation of the number of sosies that our implementation generates per hour (based on compilation and test times of table \ref{tab:data}).

\subsubsection{RQ1. Do previous results on creating sosies in imperative 
programs hold for our new tailored transformations on object-oriented 
programs?}
\label{rq1} 

Schulte et al. \cite{Schulte13} have shown the existence of sosies in the context of imperative C code.
Our experiments confirm this fact with many experimental variables that are changed.
First, our experiments are on Java, which is a different programming language, object-oriented, with richer data types and stronger typing. 
Second, our dataset covers different application domains.
Third, our transformations are more sophisticated.
Our results are thus a semi-replication.
On the one hand, we confirm the results of Schulte et al. \cite{Schulte13} on the same kind of experiment.
On the other hand, we show that sosies exist in very large quantities in a different context (different language, different dataset). We have synthesized a total of 
\nbsosies 
sosies over all programs and all kinds of transformation. As particular examples, we notice 6072 sosies for Jbehave or 1287 sosies for EasyMock with \transfo{Steroid} transformations (\transfo{add-Steroid}, \transfo{replace-Steroid}). Globally, table \ref{tab:sosies-global} shows that we synthesized sosies for all programs and with any type of transformation. Even if the quantities of sosies largely vary depending on the programs and transformations, the numbers are not in the dozens but in the hundreds, except for dagger.
This reassures us on the adequacy of software sosies for controlled unpredictability in the context of moving target defense for object-orient programs.

\subsubsection{RQ2. What are the  code transformations that confine the densest spaces for sosiefication?}
\label{rq2}

Column ``candidate'' of Table \ref{tab:sosies-global} gives the size of search space associated with each transformation (abstracting over the search space of transplantation points: we consider the same set of transplantation points for all transformations). 
Hence, by dividing the number of synthesized sosies by the number of explored candidates (column ``variant''), we obtain an approximation of the density of sosies within this search space.
This density is the key metric\footnote{we checked the sensitivity of our analysis by cutting the set of compilable variants in 10. All sets  had equivalent distributions of compilation ratios and sosie density according to a $\chi^2$ test, with a statistical significance $p<10^{-9}$ } for answering our research question.

We first analyze the problem according to the type of analysis,  i.e. whether the strategies \transfo{Rand}, \transfo{Reaction}, \transfo{Wittgenstein}, \transfo{Steroid} are similarly efficient or not. Recall that strategies \transfo{Rand} are our baseline and strategies \transfo{Steroid} are our champions.

Compilation rates for the baseline transformations, \transfo{add-Rand}, \transfo{replace-Rand}, are low (most of the generated variants do not even compile), also, most of the compilable variants are not sosies. 
They set the baseline density at approximately 10\%.

Adding analysis over variable names (\transfo{Wittgenstein}) or variable types (\transfo{Reaction}) immediately improves both compilation and density: respectively 39\% and 36\% increase in compilation rate, and 4.8\% and 3.8\% increase in density. 
The empirical results show that the matching of variables names (\transfo{Wittgenstein}) makes sense.
This indicates that the variables names carry meaning
Our results show that transplanting a variable with the same name. at different place in the programs tends to preserves the compilation and execution semantics of the program. This confirms previous results on name-based program analysis \cite{Host2009,pradel2011detecting}.
Interestingly, the sosie density of \transfo{Wittgenstein} against \transfo{Steroid} (our champions) is not that different. This is an important result: it shows that it should be possible to efficiently create sosies in dynamic languages.

The transformations on \transfo{Steroid} use both the type-based reactions and a mapping of variable names. 
\transfo{Steroid} transformations (\transfo{add-Steroid} and \transfo{replace-Steroid}) give in average the best results both in terms of compilation rates and density. 
The density of the search space of \transfo{add-Steroid} and \transfo{replace-Steroid} is higher than the density of the baseline transformations \transfo{add-Rand}, \transfo{replace-Rand} and \transfo{delete}. For instance, for JUnit, the density of \transfo{add-Steroid} is 24\% while the density of \transfo{add-Rand} is 8\%. This can be rephrased: the likelihood of finding a sosie increases from 8\% to 24\%.

Yet there is a noticeable difference between \transfo{add-Steroid} and \transfo{replace-Steroid} on all programs. This is not only for \transfo{Steroid}, for all types of analysis (\transfo{Steroid}, \transfo{Reaction}, etc), we observe a similar trend: \transfo{add} creates denser search space compare to \transfo{replace}. We explain it by the effect size of a replace: it is conceptually one delete and one add, which means that the transformation combines the behavioral effects of both. Consequently, it is less likely that those stacked behavioral changes are considered equivalent with respect to the expected behavior encoded in the test suite. 

Let us now analyze the results under the perspective of the family of transformations (\transfo{delete}, \transfo{add}, \transfo{replace}).
\transfo{delete}, which straight forward and purely random, generates a large quantity of variants that compile as well as high quantity of sosies compared to random transformations, both in absolute numbers and rates (between 5\% and 10\% of the variants synthesized with \transfo{delete} are sosies). 
This means that there is a lot of redundant or optional code in tested statements (recall that we only delete statements that are executed at least by one test case).
When \transfo{add} and \transfo{replace} transformations produce compilable variants, they are more often sosies: the search space is denser. This is especially true for \transfo{add}, which globally achieves the highest sosie density. This can be explained by the fact that the  behavior added by the transplant is outside of the expected behavioral envelop defined by the test suite.

 \transfo{Reaction}, \transfo{Wittgenstein} and \transfo{Steroid} transformations select candidates based on some definition of  compatibility (type or name-based). Consequently, the number of transplant candidates at each transplantation point is much smaller than for \transfo{rand}, which can use any statement in the program as a transplant. 
For instance, over all 669 tested transplantation points, the search space of \transfo{Steroid} consists of 7754 potential variants, which is much smaller than the two millions ($1949466$) candidates for \transfo{add-Rand}.
Recall that there are two levels of sampling: on the transplantation points, and on the transplants; to some extent, there are two nested search space.
Our results show that program analysis (\transfo{Wittgenstein},\transfo{Reaction},\transfo{Steroid}) drastically reduces the size of the transplant search space as shown by column $candidate$ of table  \ref{tab:sosies-global}. This is probably the key factor behind the increase in density. It is interesting to notice the exceptions of collections and maths, which have huge sets of candidates. These exceptions occur because of a couple of statements with very large input contexts. For example, in commons.collections we found a statement with an input context of 20 variables with the following types: $[int\times 9, strMatcher\times 3, char\times 2, boolean\times 2, strBuilder\times 1, char\times 1, List\times 1, StrSubstitutor\times 1]$ that was replaced by a reaction: 
$[int\times 6, strMatcher\times 3, char\times 2, boolean\times 2, strBuilder\times 1, char\times 1, List\times 1, StrSubstitutor\times 1]$, leading to $9^6*3^3*2^2*2^2*1^1*1^1*1^1*1^1 = 4.4*10^9$ candidates for a single transplant.

A consequence of our budget based approach is that, for small programs, the search space is small enough to allow an almost exhaustive search. For instance, in project Dagger, for all \transfo{Wittgenstein}, \transfo{Reaction} and \transfo{Steroid transformations}, we have tested between 85\% and 95\% of transplant candidates on 89.5\% of all statements that can be transformed.

The last column ``sosies/h'' (per hour), represents the sosiefication speed. 
The sosiefication speed is for a given machine, the sum of the time spent to generate \# variants with transformations, the time spent to compile them, and the time spent to run the test suite on the compilable ones, everything divided by the number of found sosies.  
This time is only indicative. There is a direct link between the density and the speed. For a given implementation and computer, if the space is denser, sosies can  be found more often.
In our implementation, the order of magnitude of the sosiefication speed is several dozens per hour. 
For instance, \transfo{add-Steroid} enables us to mine 85 sosies per hour in average for JUnit.
With respect to this evaluation criterion, \transfo{Steroid} transformations are the fastest, going up to 92 sosies per hour for \transfo{add-Steroid} on EasyMock. 
\emph{Following our metaphor, there is a boosting effect of our transformation steroids on the sosiefication speed.}

\subsubsection{RQ3. Are sosies computationally diverse, i.e. exhibit executions that are different from the original program?}

To answer RQ3, we monitor the execution of test cases on sosies as explained in Section \ref{sec:protocol}. 
We perform a pilot experiment on a sample of all sosies synthesized for Dagger, Easymock and JUnit (independently of the  transformation that is used). The random sampling ensures that the sample contains sosies generated with any of transformation.
We only consider three projects for sake of time before the deadline.
The number of sosies in each sample is in the second column of Table \ref{tab:diffTrace}. We ran the test suites on all these sosies and observed 21,255,821, 48,382 and 989,152 method calls as well as 140,902, 13,300 and 70,9113 data points for Junit, Dagger and EasyMock respectively.

Table \ref{tab:diffTrace} gives the results of this pilot experiment.
It gives the percentage of sosies on which we observe a call diversity or a variable diversity (as explained in Section \ref{sec:protocol}).
This data indicates that there are indeed a large quantity of sosies that exhibit differences in computation: 67\%, 47\%, 46\% of sosies in Dagger, Easymock and JUnit respectively exhibit at least one difference in data or method calls, compared to the computation of the original program. We also notice a great disparity in the nature of diversity between the programs. While a vast majority of the computationally diverse sosies of Dagger vary on method calls, they are much more balanced between method calls and data for Easymock and JUnit. The last two columns of the table give the mean number of test cases for which we observe a difference. These data indicate that computation diversity is not isolated and can be very important (as is the case for JUnit).

What about the sosies for which we do not observe any computation diversity? 
We see two possible answers.
First, our first implementation does not monitor everything single bits of execution data, we only focus on two specific monitoring points. If the execution difference lies somewhere else, we do not see it.
Second, those sosies might be useless sosies. For instance, a sosie whose only difference is to print a message to the console might be irrelevant in many cases. If we are not capable of assessing their computational diversity, they are not likely to hinder the execution predictability for an attacker.

\subsection{Threats to Validity}
\label{sec:threats}

We performed a large scale experiment in a relatively unexplored domain.
We now present the threats to the validity of our findings. 
First, the quality of the test suite of each program has a major impact on our findings. The more precise the test suite is (large number of relevant test scenarios and data, and as many assertions as needed to model the expected properties), the more meaningful the sosies are.
To our knowledge, characterizing the quality of assertions in test cases (\textit{i.e.} qualifying how well a test suite expresses the expected behavior) is still an open question
To mitigate this threat, we did our best to select programs for which the test suite was known to be strong in terms of coverage or reputation (the Apache foundation, which hosts all the commons libraries, has very strong rules about code quality).

Second, our findings might not generalize to all types of applications. We selected frameworks and libraries because of their popularity. 
Again, we are contributing to explore a new domain (computationally diverse sosie programs), and further experiments are required to 
confirm our findings and extend them with other application domains, programming languages (loosely-typed languages might perform differently), and computing platforms.

The last threat lies in our experimental framework. We have built a tool for program transformation and relied on the Grid5000 infrastructure to run  millions of transformations. We did extensive testing of our code transformation infrastructure, built on top of the Spoon framework that has been developed, tested and maintained for over more than 10 years.
However, as for any large scale experimental infrastructure, there are surely bugs in this software. We hope that they  only change marginal quantitative things, and not the qualitative essence of our findings. Our infrastructure is publicly available at \url{http://bit.ly/LQJYFA}.

Finally, to further reassure us on the meaningfulness of sosies, we ran the test suites of  JFreechart, PMD and commons\-math on a sample of 100 sosies of JUnit. In other terms, we applied moving target to the testing infrastructure itself.
The test suites ran correctly in 80\% of the cases. 
For the reader who would like to run his own test suites using one sosie of JUnit, they are available for download at
 \url{http://diversify-project.eu/junit-sosies/}.

\section{Related Work}
\label{sec:rw}

Mutational robustness \cite{Schulte13} 
is closely related to sosie synthesis. 
Schulte et al say that software is robust to mutations, we say that there exists transformations that introduce valuable computational diversity. 
While Schulte et al. use only random operations, we explore several types of analysis and their impact on the probability of sosie synthesis. While, Schulte et al. evaluate the effect of computation diversity to proactively repair bugs present in the original program, we provide a first quantitative evaluation about the presence of computational diversity.

Jiang et al. \cite{jiang09} identify semantically equivalent code fragments, based on input/output equivalence. They automatically extract code fragments from a program and generate random inputs to identify the fragments that provide the same outputs. Similarly to our approach, program semantics is defined through testing: random input test data for Jiang et al., test scenarios and assertions in our case. However, we synthesize the equivalent variants, and quantify the computational diversity, while Jiang et al. look for naturally equivalent fragments and do not characterize their diversity. 

More generally, sosie synthesis is related to automatic generation of software diversity. Since the early work by Forrest et al. \cite{forrest97},  advocating the increase of large-scale diversity in software, many researchers have explored automatic software diversification.

\textit{System Randomization}. A large number of randomization techniques at the systems level aim at defeating attacks such as code injection, buffer overflow or heap overflow \cite{xu03, shacham04}. The main idea is that random changes from one machine to the other, or from one program load to the other, reduces the predictability and vulnerability of programs. For instance, instruction set randomization  \cite{kc03,barrantes05} generates process-specific randomized instruction sets so the attacker cannot predict the language in which injected code should be written. 
Lin et al. \cite{lin2009}  randomize the data structure layout of a program with the objective of generating diverse binaries that are semantically equivalent. 
While this previous work focuses on transforming a program's execution environment while preserving semantic equivalence, we work on transforming the program's source code, looking for computation diversity.

\textit{Application-level diversity}. Several authors have tackled automatic diversification of application code. Feldt \cite{feldt98} has successfully experimented with genetic programming to automatically diversify controllers, and managed to demonstrate failure diversity among variants. Foster and Somayaji \cite{foster10} have developed a genetic algorithm that recombines binary files of different programs in order to create new programs that expose new feature combinations. From a security perspective, Cox et al. \cite{cox06} propose the N-variant framework that executes a set of automatically diversified programs on the same inputs, monitoring behavioral divergences. Franz \cite{franz10} proposes to adapt compilers for the automatic generation of massive scale software diversity in binaries.
None of these papers analyze the cost (in terms of trial, search space size or time) of diversity synthesis and only Feldt explicitly targets computation diversity.

\textit{Unsound transformations}. There is a recent research thread on so-called unsound program transformations \cite{rinard11}. 
Failure-oblivious computing \cite{rinard04} consists in monitoring invalid memory accesses, and crafting return values instead of crashing, letting the server continue its execution. Following this idea, loop perforation  \cite{misailovic10} monitors execution time on specific loops and starts skipping  iterations, when time goes above a predefined threshold.
Automatic program repair \cite{monperrus14}  also relies on program transformations with no semantic guarantees. For example, Le Goues et al. propose an evolutionary technique to transform a program that has a bug characterized by one failing test case into a program for which this test case passes \cite{legoues12}. Carzaniga et al. \cite{carzaniga13} have a technique for runtime failure recovery based on diverse usages of a faulty software component.
Sosiefication exactly goes along this line of research. The code transformations might also introduce differences in behaviors that are not specified.

\textit{Natural diversity}. Sosie synthesis is  about artificial, automated software diversity. There is also some ``natural software diversity''.
For example, there exists a diversity of open source operating systems (that can be used for fault tolerance \cite{koopman99}) or a diversity of virtual machines, useful for moving target defense \cite{christodorescu11}. Component-based software design is an option for letting a natural resuable diversity of off-the-shelf components \cite{totel06}.
This natural diversity comes from both the market of commercial competitive software solutions and the creativity of the open-source world \cite{hertel03}.

\section{Conclusion}
\label{sec:conclusion}

We have explored the efficiency of 9 program transformations that add, delete or replace source code statements, for the automatic synthesis of \emph{sosies}, program variants that exhibit the same behavior but different computation. We experimented sosie synthesis over \nbprograms Java programs and observed the existence of large quantities of sosies. In total, we were able to synthesize \nbsosies sosies. We observed that considering type and variable compatibility is the most efficient in terms of absolute number of sosies and sosiefication speed.

We consider this work as an initial step towards controlled and massive unpredictability of software. Next steps include two aspects related to the effectiveness of our process. 

In the context of moving target defense, one can think of generating sosies on the fly.
Thus, one would set requirements on the number of variants to be used, on the number of behavioral moves required to keep attacks hard.
Let us assume that one needs to have 10 new variants every hour. 
In this case, one needs a process that generates at least 10 new variants per hour. Furthermore, there is not only a constraint on the number of new sosies, but also a constraint on the novelty, i.e. the dose of unpredictability that each sosie brings. In this context, novelty search seems to be a good technique. In this line, the key question we would like to answer is: 
is there an upper limit on the number of sosies one could create practically, or is 
computational diversity unbounded?

\newpage 
\bibliographystyle{abbrv}
\bibliography{biblio}

\end{document}